\documentclass[final]{aipproc}

\layoutstyle{6x9}

\begin{document}

\title{The meson spectra beyond a $q\bar q$ description}

\author{J. Vijande, F. Fern\'andez and A. Valcarce}
{address={Nuclear Physics Group, University of Salamanca, 
E-37008 Salamanca, Spain}}

\begin{abstract}
Despite the apparent simplicity of meson spectroscopy there are some states
which cannot be accommodated in the usual $q\bar q$ structure. Among
them there are either exotic states as the $X(1600)$, or 
the recently measured charmed states $D_{sJ}^*$ and
some of the light scalar mesons. We present a
description of the light scalar mesons in terms
of $q\bar q$ and $qq\bar q\bar q$ components. 
\end{abstract}

\maketitle

Nearly all known mesons made of $u$, $d$, and $s$ quarks fit neatly
into the multiplets expected in generic constituent quark models. 
The single, striking exception were the scalar mesons, i.e., $J^{PC}=0^{++}$.
Either they form an anomalously light nonet or 
a nonet and more scalar mesons appear 
in the energy region 1.2$-$1.5 GeV \cite{Ja04}. 
Such a long-standing situation has been
re-invigorated by the new data obtained at BaBar, CLEO, FOCUS
and Belle, creating a challenging scenario. There have been
reported a number of states compatible with meson quantum numbers, 
as the $D^*_{sJ}$'s \cite{BC03}, but either their masses do not fit
into the quark model predictions in its many variations or they over-populate
the expected number of states. For years multiquark states have
been justified to coexist with $q\bar q$ states in the energy region
around 1 GeV for the case of the light mesons \cite{Ja77}. This situation 
claims for a comprehensive study where two- and four-quark states are
considered simultaneously. 

A $qq\bar q \bar q$ state has the same quantum numbers and the same
quark content as a $q\bar q - q \bar q$ meson scattering state. In a 
fairly precise way the $qq \bar q \bar q$ state can be considered as 
a piece of the meson-meson continuum that has been artificially 
confined by a confining boundary condition or potential that is
inappropriate in the meson-meson channel \cite{Ja79}. If the 
multiquark state is unusually light or sequestered (by the spin,
color and/or flavor structure of the wave function) from the
scattering channel it may be prominent. If not, it would be
just a silly way of enumerating the states in the continuum.

In this work we present a careful study of possible 
prominent four-quark states in the low-energy meson spectra.
For this purpose, we will address the description of hadrons with 
zero baryon number described as clusters of quarks confined
by a realistic interaction between them (it allows to
describe the NN data and the baryon spectra).
Our work tries to supersede other studies of four-quark systems
devoted either to a particular set of states \cite{Te03}, or 
more general studies that in any case made a detailed
comparison with $q\bar q$ predictions within the same model \cite{ZS86}.

We have solved the two-body problem by means of the 
Schr\"odinger equation treating in an exact way the non-central
terms of the interacting potential.
The four-body problem has been solved by means of 
a variational method using as trial wave function 
the most general linear combinations of gaussians \cite{SV98}. 
In particular, the non-quadratic terms that were neglected 
in previous works \cite{VF03c}, have been considered. While 
they are known to play a minor role in the description of 
light-heavy tetraquarks, they have a great influence in the light case.
The mixing between the two- and four-body components will be 
parametrized as explained in the following.

The interacting potential and the method to solve the
four-body problem have been tested in the case of a system
whose quantum numbers can only be obtained by means 
of a multiquark description, the 
isospin two $X(1600)$. This state has been 
observed in the reaction $\gamma\gamma\rightarrow\rho\rho$ near
threshold, reported with a mass of $1600\pm100$ MeV and
quantum numbers $I^GJ^{PC}=2^+(2^{++})$ \cite{PDG02}. 
It cannot be described as 
a $q\bar q$ state, being therefore an exotic meson. 
It can be easily understood as 
a tetraquark made of four light quarks coupled to $I=2$,
$S=2$ and $L=0$. Our formalism predicts for this 
configuration an energy of 1500 MeV,
in excellent agreement with the experimental data, giving 
confidence to the results obtained for the four-body configurations.

In non-relativistic quark models gluon degrees 
of freedom are frozen and therefore the total wave 
function for a $B=0$ hadron may be written symbolically as
\begin{equation}
\label{mes-w}
\left|\rm{B=0}\right>=\sum_n\Omega_n\left|(q\bar q)^n\right>=
\Omega_1\left|q\bar q\right>+\Omega_2\left|q\bar q q\bar q\right>+....
\end{equation}
where $q$ stands for quark degrees of freedom.
As mentioned above, the energy of the $q\bar q$ and $qq\bar q\bar q$ 
configurations can be obtained from our hamiltonian, the results being
resumed in table \ref{mix2}. One can see that neither the
$q \bar q$ nor the $qq\bar q \bar q$ configurations match
the experimental data. To consider the mixing between the two- and
four-body configurations (the calculation of the coefficients $\Omega_i$), 
requires the knowledge of the operator annihilating a quark-antiquark pair 
into the vacuum. This could be done, for example, using a $^3P_0$ model,
but the result will always depend on the parameters used to describe
the vertex. Therefore we have decided to parametrize these coefficients
by looking to the quark pair that it is annihilated, and
not on the spectator quarks that will form the final $q\bar q$ state. 
\begin{eqnarray}
\left<qq\bar q\bar q|O|q\bar q\right>=\left<qs\bar q\bar s|O|s\bar s\right>
=\left<qq\bar q\bar s|O|q\bar s\right>&=&C_q\\ \nonumber
\left<ss\bar s\bar s|O|s\bar s\right>=\left<qs\bar q\bar s|O|q\bar q\right>
=\left<qs\bar s\bar s|O|q\bar s\right>&=&C_s.
\end{eqnarray}
The mixing parameters, $C_s$ and $C_q$, have been fixed using
the best measured scalar states: the $a_0(980)$ and the $f_0(980)$. 

\begin{table}
\caption{Mass, in MeV, of the $q\bar q$ and $qq\bar q \bar q$
light isoscalar, isovector and I=1/2 systems.}
\label{mix2}
\begin{tabular}{ccc|ccc|ccc}
\hline
\multicolumn{3}{c}{I=0} & 
\multicolumn{3}{|c|}{I=1} & 
\multicolumn{3}{c}{I=1/2}  \\ 
Meson & $q \bar q$ & $qq\bar q \bar q$  &
Meson & $q \bar q$ & $qq\bar q \bar q$  & 
Meson & $q \bar q$ & $qq\bar q \bar s$  \\
\hline
$f_0(600)$ & 648  & $-$  &
$a_0(980)$ & 1079 & $-$ &
$\kappa(800)$ & 1273 & $-$ \\
$f_0(980)$ & $-$  & 949  &
$a_0(1450)$ & $-$ & 1308  &
$K_0^*(1430)$ & $-$ & 1295 \\
\hline
\end{tabular}
\end{table}

\begin{table}
\caption{Mass, in MeV, and flavor dominant component for the 
light isoscalar, isovector and I=1/2  mesons.}
\label{mix1}
\begin{tabular}{ccc|ccc|ccc}
\hline
\multicolumn{3}{c}{I=0} & 
\multicolumn{3}{|c|}{I=1} & 
\multicolumn{3}{c}{I=1/2}  \\ 
Mass & Exp. & Flavor &
Mass & Exp. & Flavor & 
Mass & Exp. & Flavor \\
\hline
568&400$-$1200&$q\bar q$ &
985&984.7$\pm$1.2&$q\bar q$ &
1113&$\approx$ 800&$q\bar s$\\
999&980$\pm$10&$qq\bar q\bar q$ &
$\begin{array}{c}1381\\1530\end{array}$&1474$\pm$19&
$\begin{array}{c} qq\bar q\bar q\\ qs\bar q\bar s \end{array}$ &
1440&1412$\pm$6&$qq\bar q\bar s$\\
$\begin{array}{c}1301\\1465\end{array}$&1200$-$1500&
$\begin{array}{c} s\bar s\\ q\bar q \end{array}$ &
1640&&$q\bar q$&
$\begin{array}{c}1784\\1831\\2060\end{array}$&1945$\pm$20&
$\begin{array}{c} q\bar s\\ qs\bar s\bar s \\q\bar s \end{array}$\\
1614&1507$\pm$5&$qs\bar q\bar s$ &
1868&&$q\bar q$ & & &\\
1782&1713$\pm$6&$q\bar q$ & & & & & & \\
$\begin{array}{c}1900\\1944\end{array}$&1992$\pm$16&
$\begin{array}{c} s\bar s\\ ss\bar s\bar s \end{array}$
& & & & & & \\
$\begin{array}{c}2224\\2351\end{array}$&2197$\pm$17&
$\begin{array}{c} s\bar s\\ s\bar s\end{array}$ & & & & & & \\
\hline
\end{tabular}
\end{table}
 
The obtained mass and dominant flavor component for all 
the scalar mesons are given in table \ref{mix1}. The
experimental data are taken from Ref. \cite{PDG02}.
One should notice that such an interpretation of the light
scalar mesons in terms of two- and four-quark components
allows for a one-to-one correspondence between theoretical
states and experiment. Our results assign a dominant tetraquark
component to the $f_0(980)$, $a_0(1450)$ and the $K_0^*(1430)$.
The final physical picture arising from these results shows 
an involved structure for the flavor wave function of the
light scalar mesons, in agreement with the complicated pattern
observed for their decays. From our results one can also
see how the predicted isoscalar state between 1.2 and 1.5 GeV
may in fact correspond to two different resonances, as suggested
by the PDG. We also observe that the description of the 
high energy states is worse that the low-energy data, what 
may be a consequence of not including in our calculation 
excited states of the four-quark configurations. 

It seems therefore
coherent to include simultaneously the two- and four-quark
components of the wave function to describe the light scalar mesons.
Preliminary studies of the charmed sector allows for a similar
correspondence between reported states and theoretical numbers.
The study of the pattern decays should definitively
conclude the correctness of the present description.

\begin{theacknowledgments}
This work has been partially funded by Ministerio
de Ciencia y Tecnolog{\'{\i}}a under Contract No. BFM2001-3563
and by Junta de Castilla y Le\'{o}n under Contract No. SA-104/04. 
\end{theacknowledgments}

\end{document}